# Direction- and Salt-Dependent Ionic Current Signatures for DNA Sensing with Asymmetric Nanopores

Kaikai Chen,[1,2] Nicholas A. W. Bell,[1,†] Jinglin Kong,[1] Yu Tian,[2] and Ulrich F. Keyser[1,*]
[1]Cavendish Laboratory, University of Cambridge, JJ Thomson Avenue, Cambridge, CB3 0HE, United Kingdom; and [2]State Key Laboratory of Tribology, Tsinghua University, Beijing 100084, China

*Correspondence: ufk20@cam.ac.uk
†Present address: Department of Chemistry, University of Oxford, Oxford, OX1 3TA, United Kingdom

ABSTRACT Solid-state nanopores are promising tools for single molecule detection of both DNA and proteins. In this study, we investigate the patterns of ionic current blockades as DNA translocates into or out of the geometric confinement of such conically shaped pores. We studied how the geometry of a nanopore affects the detected ionic current signal of a translocating DNA molecule over a wide range of salt concentration. The blockade level in the ionic current depends on the translocation direction at a high salt concentration, and at lower salt concentrations we find a non-intuitive ionic current decrease and increase within each single event for the DNA translocations exiting from confinement. We use recently developed DNA rulers with markers and finite element calculations to explain our observations. Our calculations explain the shapes of the signals observed at all salt concentrations and show that the unexpected current decrease and increase are due to the competing effects of ion concentration polarization and geometric exclusion of ions. Our analysis shows that over a wide range of geometry, voltage and salt concentrations we are able to understand the ionic current signals of DNA in asymmetric nanopores enabling signal optimization in molecular sensing applications.

## INTRODUCTION

Nanopores have become a powerful technique for single molecule sensing of polynucleotides such as DNA and RNA (1). As an alternative to biological pores such as alpha-hemolysin, solid-state nanopores fabricated with silicon nitride, glass, graphene or molybdenum disulfide membranes have found versatile applications in single molecule detection and for the study of confined transport (2-11). The basic principle of nanopore sensing is that a single molecule can be detected by measuring the ionic current change during its translocation through the pore, named as the resistive-pulse method (12). It is intuitive that the current drops during the translocation since the molecule increases the pore resistance by blocking the space which would otherwise be filled with electrolyte ions. Indeed such reductions have been consistently observed at high salt concentrations. However at low salt concentrations, current increases and multi-level current blockades have been observed due to translocating particles or molecules in a variety of nanopore and nanochannel geometries (13-21). As the salt concentration decreases, the Debye length becomes longer and may be comparable to the pore size, thus surface charge begins to play a more important role (22). In this case the ionic current change is determined not only by the geometric exclusion of ions but also the change in ion concentration due to the translocation molecule or particle (16-19).

Asymmetric conical nanopores pulled from quartz or borosilicate capillaries are frequently used in single molecule detection, with the advantages of simple and inexpensive fabrication, low noise and multiplexed measurement capabilities (8, 23). These conically shaped glass nanopores have been used for the detection of a range of biomolecules such as DNA (24-26) and proteins (27-29). In all translocation experiments we have reported so far, the negatively charged biomolecules move into the conical nanopore from a large reservoir outside, i.e. a positive potential was applied inside the conical nanopore. With low salt concentration, the conical shape of the pore causes current rectification and also electroosmotic flow rectification (30,31). It is important to study the underlying physics behind DNA translocations and fully understand the current signatures at a variety of experimental conditions because this provides more options for biological molecule sensing such as salt concentrations close to the physiological environment.

In this report, we studied DNA translocations through ~15 nm diameter glass nanopores with salt concentrations from 4 M to 20 mM. We examine the ionic current signatures caused by the DNA entering or exiting the conical nanopore confinement. The asymmetric pore geometry gives rise to a variety of distinct ionic current signals depending on the salt concentration and translocation direction. At high salt concentrations exemplified by 4M LiCl, a current reduction is exclusively observed but with a small difference in magnitude between the two translocation directions. At ~1M LiCl or KCl there is a distinctive difference between translocations into and out of the conical confinement. Translocations into confinement show a uniform current blockade but translocations out of confinement create a biphasic pattern, i. e. current decrease and increase in a single event. We use finite element calculations to simulate the distribution of ions in the pore and the current change caused by a charged rod representing the DNA. Our model reveals that the ion concentration is modulated within the nanopore as the DNA passes the pore. The current change is ultimately caused by a combination of electrolyte concentration modulation and geometric exclusion of ions by the DNA. Our simulations account for the shapes of the observed ionic current blockades and their dependence on salt concentration.

**MATERIALS AND METHODS**

**Glass nanopore fabrication**

Glass nanopores were pulled from quartz capillaries with outer diameter 0.5 mm and inner diameter 0.2 mm using a commercial pipette puller (P2000, Sutter Instruments), with the same parameters shown in ref 27 where the final tip diameter was estimated at 15±3 nm (mean±s.d.) from scanning electron microscopy.

**DNA samples and chemicals**

Double-stranded DNA samples (NoLimits DNA fragment) were purchased from Fisher Scientific with the following lengths: 3, 5, 7, 8, 10 and 15 kbp. The DNA ruler was synthesized according to the method shown in ref 38. LiCl and KCl powders were purchased from Sigma-Aldrich. Solutions with different salt concentrations were all buffered with 1×Tris-EDTA (Sigma-Aldrich, 10 mM Tris, 1 mM EDTA), except for the 20 mM KCl and LiCl solutions for which 0.2×Tris-EDTA was used. The pH values were adjusted to 8 for all solutions using HCl/LiOH or HCl/KOH solutions.

**Setup and ionic current measurement**

The ionic current was recorded by an amplifier Axopatch 200B (Molecular Devices) with the current signal filtered at 50 kHz and then collected using a data card (PCI 6251, National Instrument) at a sampling frequency of 250 kHz. The data writing, voltage control and data analysis were performed with custom-written Labview programs.

**Forward and backward translocation measurement**

Forward and backward translocations were achieved by switching the voltage between a positive value and the opposite one with a period of 60 s, keeping the voltage constant at each polarity for 30 s. The DNA solutions were added in both reservoirs for the data recording and only on one side for the verification of translocation direction.

**RESULTS AND DISCUSSION**

**Ionic current traces and current change during translocation**

Nanopores with orifice diameters 15±3 nm were pulled from quartz capillaries (see methods). The bulk reservoir outside the nanopore tip was grounded as shown in Fig. 1 *A* and the DNA molecules moved into the conical nanopore at a positive voltage and out of the conical nanopore at a negative voltage. We define these translocation directions as "forward" and "backward" respectively (Fig. 1 *A*). We use two types of salts - KCl and LiCl. KCl has been the most widely used electrolyte in nanopore experiments due to the similar ion mobilities of anions and cations. LiCl solution was recently shown to slow down DNA translocations which results in a significant increase in the sensing resolution due to reduced translocation velocities (32).

In the analysis, the current change $\Delta I$ is defined as $|I|-|I_{base}|$ ($I$ is the real-time ionic current and $I_{base}$ is the base current), which is positive/negative if the current increases/decreases relative to the baseline. Fig. 1 *C* shows a summary of typical current traces with LiCl solutions and exemplary, so-called unfolded events which are caused by the linear threading of the DNA through the pore without folds or knots (33). At a concentration of 4 M, current decreases were observed during translocations in either direction (Fig. 1 *C*). In 1 M LiCl solution, the current decreased during the forward translocation but for the backward translocation, a biphasic current change appeared with a current decrease first and then an increase. When the concentration of LiCl decreased to 20 mM, as shown in Fig. 1 *C*, the current change for backward translocation turned to a sole increase, as previously observed for conical pores at low salt concentrations (34).

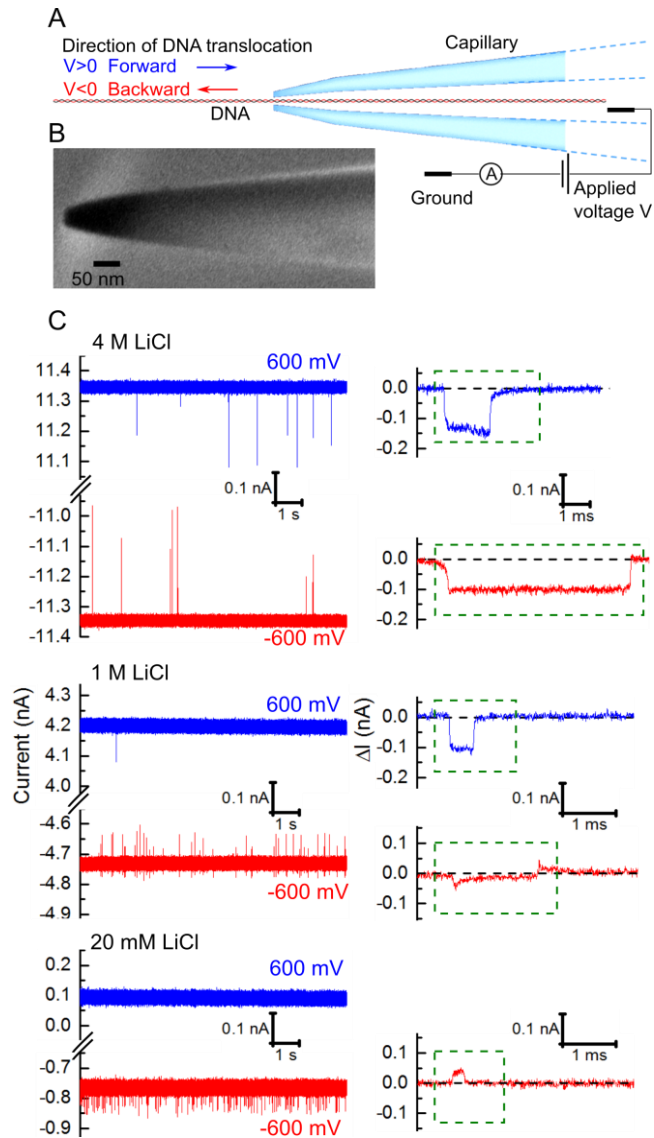

FIGURE 1   Schematic of DNA translocation and examples of ionic current traces. (*A*) Simplified diagram of DNA translocation through a conical glass nanopore with the voltage applied inside the conical nanopore. (*B*) Scanning electron microscope (SEM) image of a typical nanopore tip showing the outer dimensions. (*C*) Raw current traces (left) and current change $\Delta I$ ($|I|-|I_{base}|$) for translocation events (right, with events marked in the green dashed rectangles) during 8 kbp DNA translocations through ~15 nm diameter pores in 4 M, 1 M and 20 mM LiCl solutions at +600 mV (blue) and -600 mV (red). I-*V* curves for the pores are shown in Fig. S1. DNA was added on both sides with concentrations of 0.95 nM in 4 M and 1 M LiCl solutions and 0.19 nM in 20 mM LiCl solution.

**Direction dependence of event depth with high salt concentration**

Forward and backward translocations of 8 kbp DNA through the ~15 nm-diameter nanopores were performed in 4 M LiCl, 2 M LiCl, 4 M KCl and 2 M KCl solutions and typical current changes for unfolded events are shown in Fig. 2 *A*. For 4 M and 2 M LiCl we observe that the magnitude of the current decrease is different for the different directions despite using the same magnitude of voltage. At 2 M KCl solution the ionic current blockade for a backward translocation shows a biphasic shape. To show the current change quantitatively, we calculated the unfolded event level ("1-level current" $\Delta I_{\text{1-level}}$) by fitting the histograms of current change for all events using Gaussian functions as the folded parts did not contribute to the fitting of the unfolded parts (Fig. S2). As shown in Fig. 2 *B*, the levels were different at +600 mV and -600 mV in 4 M LiCl solution (histograms at 400~800 mV and -400~-800 mV are shown in Fig. S2). Since there was a slight slope for the current blockade level for conical pores, we studied the length dependence of the event shape and 1-level current (Fig. S3). Our results show that the direction-dependent event depth was more significant for longer DNA, and the derived $\Delta I_{\text{1-level}}$ did not change significantly further when the length was above 7 kbp, so the slight slope did not have a significant effect on the derived 1-level current for the 8 kbp DNA. The voltage dependence of the current blockades for forward and backward translocations is shown in Fig. S2 and the difference was more obvious at higher voltage. This illustrates that even in 4 M LiCl solution, with a Debye length of ~0.15 nm which is much smaller than the pore diameter ~15 nm, the event depth shows a direction dependence. Similar direction-dependent event depth profiles have been seen for particles of diameter above 100 nm translocating asymmetric nanopores with final tip diameters of several hundred nanometers in salt concentrations (100 mM or 10 mM) where the Debye length is much less than the pore diameter (19,35).

The voltage dependence of the normalized current change in both directions in 4 M LiCl and 2 M LiCl solutions are shown in Fig. 2 *C* and Fig. 2 *D*. Results for more pores in Fig. S4 and Fig. S5 also show directional dependence of the current change but with slight differences in the voltage-dependence of the normalized current change. As the LiCl concentration went down from 4 M to 2 M, the Debye length increased and also the effective surface charge on the DNA increased (32), both enhancing the effect of ion concentration polarization - a well-known property of nanochannels resulting from the preferred transfer of one ion over another and causing local ion concentration depletion or enrichment of co- and counterions, respectively (36). Experiment for particles showed the trends were not the same for different geometries and surface charge states, and the voltage dependence was sometimes even non-monotonic (35), in accordance with our experimental results for nanopores and DNA.

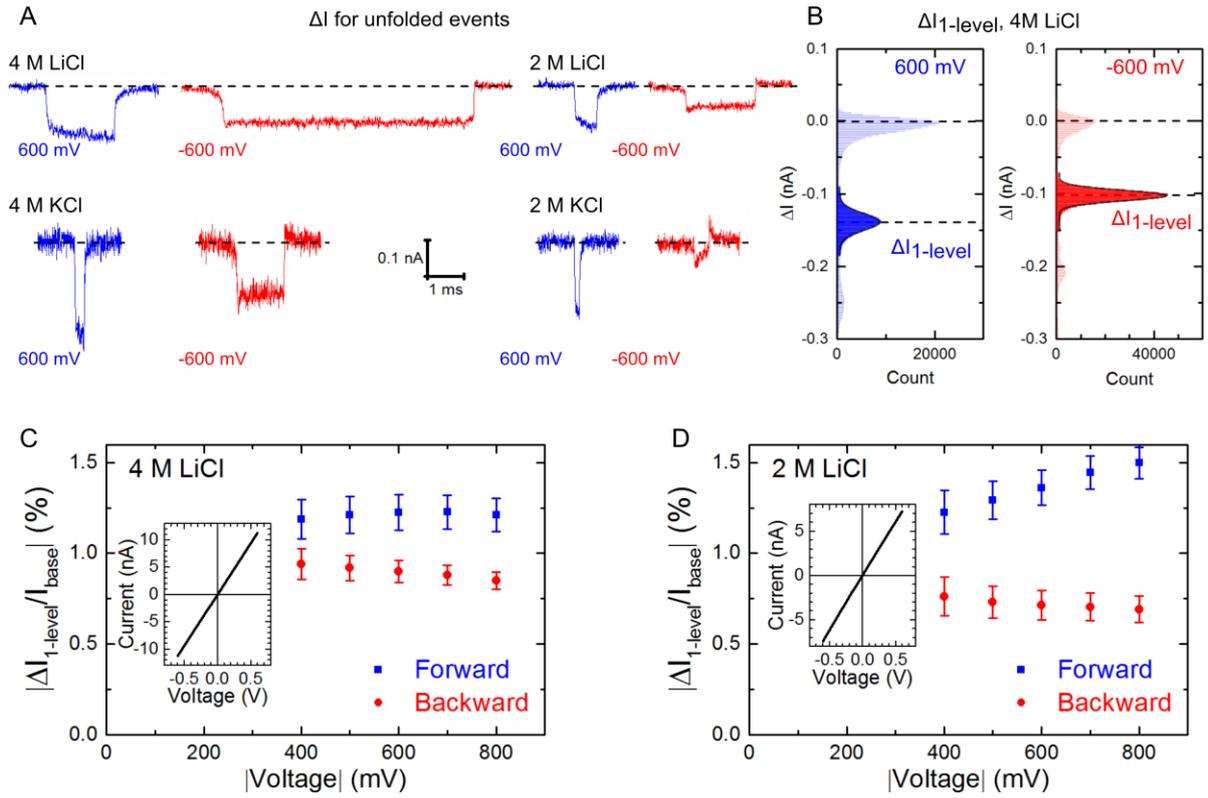

FIGURE 2  Event depths for forward and backward translocations in 2 and 4 M solutions. (*A*) Typical current changes for the unfolded events during 8 kbp DNA translocations in 4 M LiCl, 2 M LiCl, 4 M KCl and 2 M KCl solutions at ±600 mV. (*B*) An example for calculating the "1-level current" $\Delta I_{\text{1-level}}$ with 491 events at 600 mV and 396 events at -600 mV in 4 M LiCl. (*C*) and (*D*) show the absolute values of normalized $\Delta I_{\text{1-level}}$ ($\Delta I_{\text{1-level}}/|I_{\text{base}}|$) during 8 kbp DNA translocations in 4 M and 2 M LiCl solutions. Error bars in (*C*) and (*D*) represent standard deviations of the Gaussian fit to the ionic current distributions.

**Direction-dependent event shapes with medium salt concentration**

For the events with high salt concentration shown above, the current decreased in both translocation directions (except for the 2 M KCl solution). Fig. 3 shows examples of the unfolded events with 1 M KCl and 1 M LiCl solutions (more events are shown in Fig. S6). For the backward events, the current decreased at the beginning, increased slowly in the middle part, and ended with a peak. The negative and positive peak amplitudes for backward translocations were much smaller than the event depths for the forward translocations at the same voltage amplitude. Similar profiles were reported for particle or polymer translocations at low salt concentrations in former reports where they are attributed to ion concentration polarization (19,20). From these findings we expect that at concentrations of 1 M the electrical double layer (EDL) plays a more important role. The current increase during backward translocation in 1M KCl solution was more significant than that in 1 M LiCl solution. Also, the current increase at the end of an event was observed with 2 M KCl solution but not with 2 M LiCl solution. This difference can be explained by the expected dependence of DNA effective surface charge densities on the type of counter ion (32). At the same molarity, the effective charge of DNA is significantly higher in KCl compared to LiCl which results in a stronger ion concentration polarization effect as the DNA translocates through. This point is discussed further in the modelling section.

Fig. 3 *B* shows the current change for unfolded events at voltages ranging from 500 mV to 800 mV and from -500 mV to -800 mV in 1 M KCl solution. The derived normalized 1-level current for forward translocations as a function of voltage is shown in Fig. 3 *C*, where the values increased monotonically with voltage for the 3 pores.

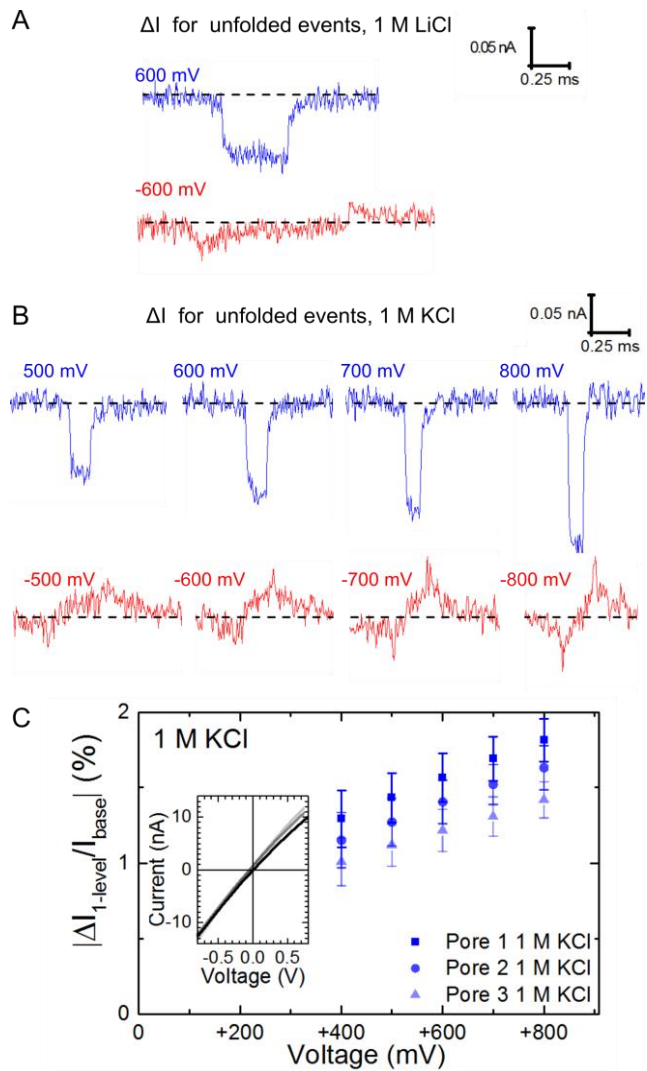

FIGURE 3  Direction-dependent event shapes for DNA translocations with 1 M salt concentration. (*A*) $\Delta I$ for unfolded events during 8 kbp DNA translocations in 1 M LiCl solution at ±600 mV. (*B*) $\Delta I$ for unfolded events during 8 kbp DNA translocations at different voltages in 1 M KCl solution. (*C*) Absolute values of normalized current change for forward translocations of 8 kbp DNA in 1 M KCl solution against voltage for 3 pores. Error bars in (*C*) show standard deviations of the Gaussian fit to the ionic current distributions.

**Current increase during backward translocations with low salt concentration**

With the salt concentration going further down to 100 mM, the current increased during the backward translocations (Fig. 4 *A*), while forward translocations were not observed for our ~15 nm pores. The absence of translocation in the forward direction is likely due to electro-osmotic flow which opposes the electrophoretic motion of the DNA. This electro-osmotic flow was extensively characterized by Laohakunakorn *et al.* and shown to increase strongly with decreasing salt concentration (31). Furthermore the flow is asymmetric with respect to voltage reversal with significantly larger flow rates measured at positive voltages. We note that for conical nanopores with larger diameters of around 50 nm, forward translocations are sometimes observed at low salt concentrations (34).

Similar to the event shapes for forward translocations with high salt concentration, one can clearly identify unfolded events, as shown in Fig. 4 *A* and Fig. 4 *B*. Using the same method shown in Fig. 2 *B*, we analyzed the normalized 1-level current for 8 kbp DNA translocations in 100 mM KCl and 20 mM KCl solutions, with the results shown in Fig. 4 *C* and Fig. 4 *D*. The normalized current change increased monotonically or showed a non-monotonic behavior, depending on the pore conductance determined by the pore diameter and conical angle. It is notable that for translocations in the 20 mM KCl solution, the normalized current rise was up to 15%, which was much higher than the normalized current drop for translocations with high salt concentration (37).

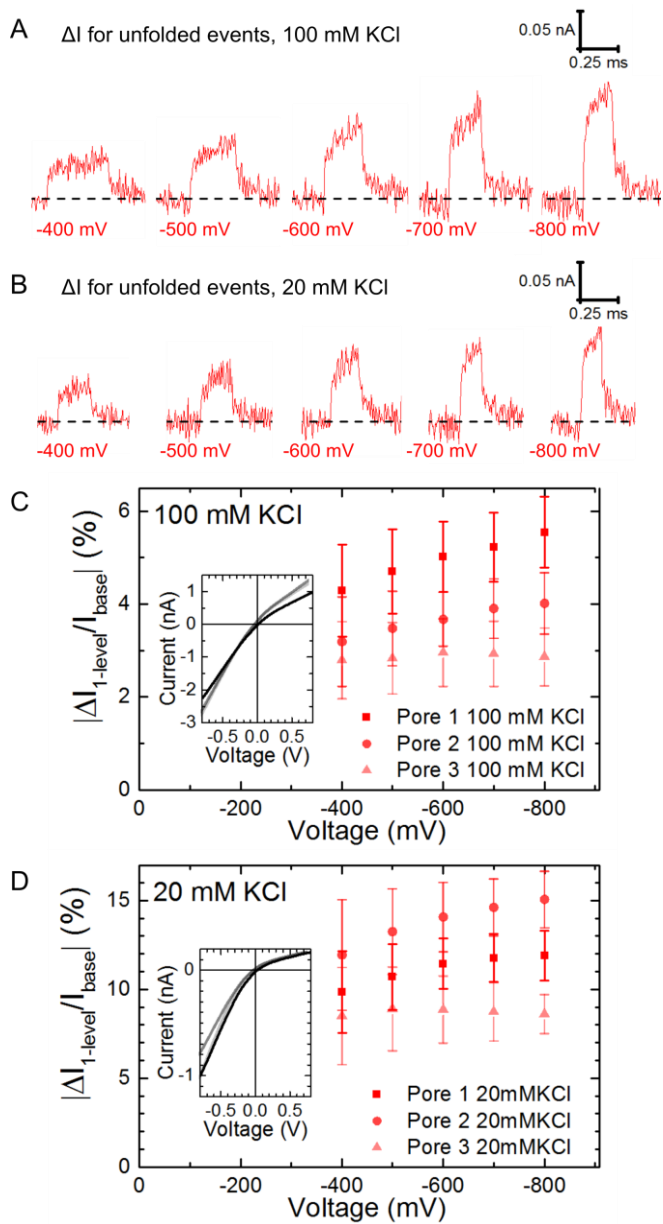

FIGURE 4 Current increase during backward translocations with 100 and 20 mM concentrations. (*A*) and (*B*) show the Δ*I* for backward translocations of 8 kbp DNA in 100 mM and 20 mM KCl solutions. (*C*) and (*D*) show the corresponding normalized 1-level current as a function of voltage with results for 3 pores at each salt concentration. Error bars in (*C*) and (*D*) show standard deviations of the Gaussian fit to the ionic current distributions.

**Translocations using DNA rulers**

In order to understand the origin of the diverse event shapes, we conducted experiments using our recently developed DNA rulers (38). The ruler consists of a 7.2 kbp backbone of double-stranded DNA with six equally spaced zones of hairpin loops protruding from the backbone, as shown in Fig. 5 *A*. The ruler is of practical importance for assessing DNA translocation and it helps to determine where the DNA is positioned in the pore when a particular ionic current level is detected (38). Fig. 5 *B* shows typical ionic current blockades of unfolded events caused by the ruler in 1 M LiCl in forward and backward directions. Previous analysis for our nanopores, based on charge exclusion, has suggested that ionic current signatures above baseline noise are caused only by full translocations (39). The DNA ruler confirms that the ionic blockade is due to the DNA fully translocating the nanopore and also allows us to relate the position of the DNA in the nanopore to the ionic blockade at a certain time during the translocation. For the backward translocation the largest current decrease occurred at approximately the same time as the first marker exits the nanopore, i. e. approximately 1/7 of the way through the translocation. The largest point of the current increase occurred after the last marker passed through indicating that the current increase is in the last section of the translocation where >6/7 of the DNA has already exited the nanopore.

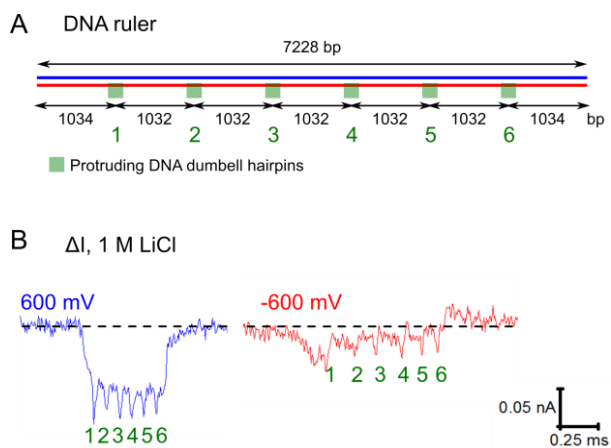

FIGURE 5 Schematic of a DNA ruler (*A*) and Δ*I* for DNA ruler translocations at ±600 mV with 1 M LiCl solution (*B*).

**Numerical simulation**

Having investigated the experimental characteristics of the ionic current blockades, we conducted finite element analysis numerical simulations to investigate the mechanisms behind the observed phenomena. Our simulation solves the Poisson−Nernst−Planck and Navier−Stokes equations (PNP-NS equations) (40) using COMSOL Multiphysics 4.4 with an axially symmetric geometry shown in FiG. S8 which is an average from nanopores measured by SEM (more model details are given in the Supporting Material). The DNA is simulated as a cylindrical rod of diameter 2.2 nm and length 2720 nm. The mesh size was refined to be 0.1 nm at the boundaries of the pore and the rod representing DNA. Fig. 6 *A* shows the calculated ion concentration profiles at both ±600 mV with 1 M KCl. For direct comparison, Fig. 6 *B* shows the profile at -600 mV with 20 mM KCl. We assume a surface charge density of -0.01 C/m$^2$ on the pore walls and -0.018 C/m$^2$ on DNA which is in the same order of magnitude as values estimated in 1 M KCl solution in the literature (41, 42). The surface charge is an important parameter in our model which is known to vary with salt type and concentration (43). The uncertainty of the surface charge prevents any quantitative analysis as well as other factors such as molecule trajectory and interaction. The latter also cause variations in the calculated current levels. While a model incorporating a surface related molecular drag would give a better quantitative explanation on the current blockade (44,45), we concentrate here on explaining the event shape rather than an absolute quantitative value. In Fig. 6 *A* and Fig. 6 *B* we systematically vary the position of the DNA thereby reflecting the passage of the DNA during a translocation. The calculated ionic current-DNA position traces are displayed in Fig. 6, *C-E*, in accordance with the experimental data shown above (note the translocation direction).

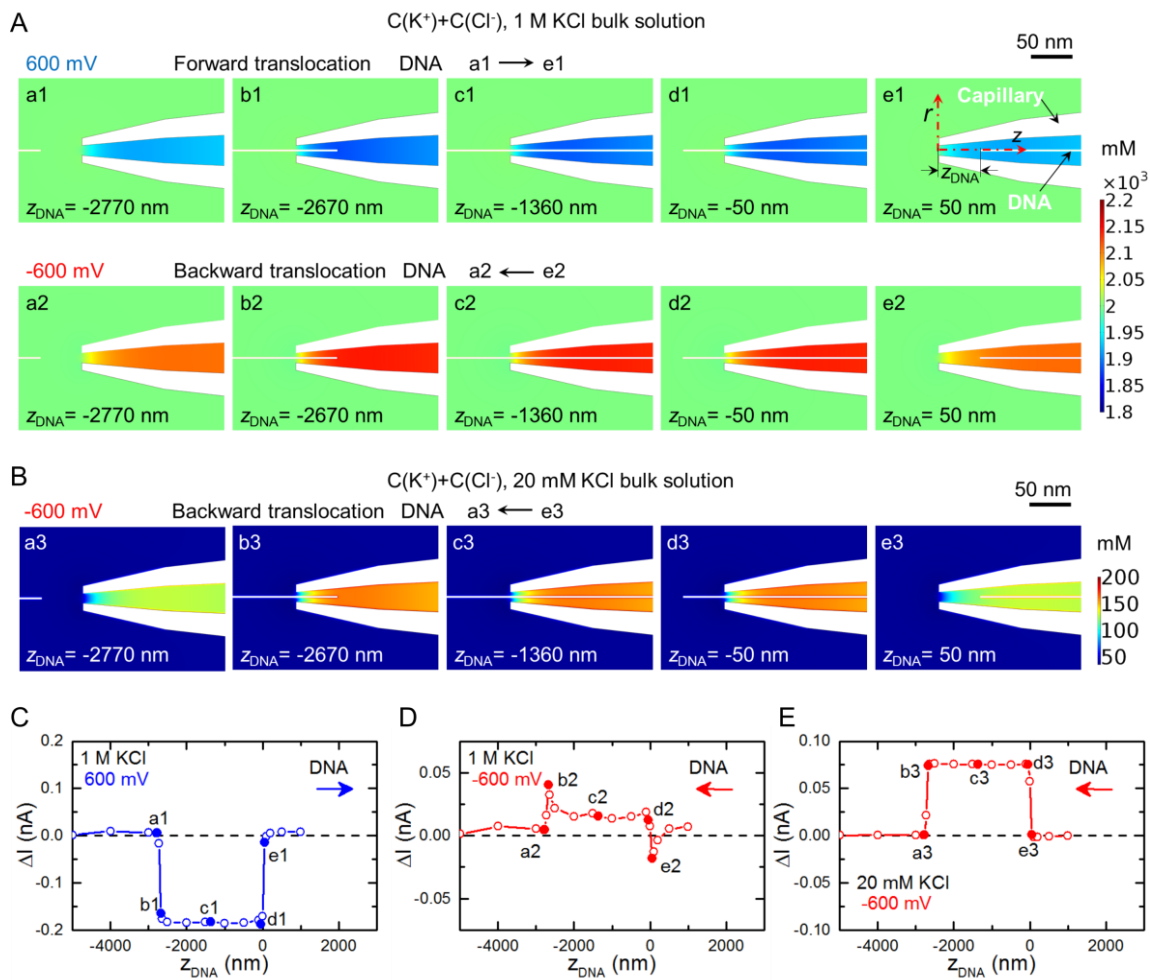

FIGURE 6 Simulation for the direction-dependent current blockades. (*A*) and (*B*) show the sum of potassium and chloride ion concentrations (C(K$^+$)+C(Cl$^-$)) with particular DNA

positions with 1 M and 20 mM bulk KCl concentrations respectively. Definition of the DNA position $z_{DNA}$ is shown in the top right corner. (*C*) and (*D*) show the current change as a function of DNA position at ±600 mV in 1 M KCl solution. (*E*) Current change as a function of DNA position at -600 mV in 20 mM KCl solution. Arrows show directions of DNA translocation. Pore surface charge density $\sigma_{pore}$ -0.01 C/m² and DNA surface charge density $\sigma_{DNA}$ -0.018 C/m² are used in the simulation.

Our simulations reveal a difference between the concentrations of ions inside the conical nanopore and those in the bulk reservoir, which is known as ion concentration polarization (36). In our experiment the selectivity arises from the negative charge of the pore walls and DNA which induce a cloud of positively charged counterions. The magnitude of ion concentration around the nanopore tip changes depending on the position of the DNA in the nanopore. There is 5 to 10 percent difference in Fig. 6 *A* and this concentration modulation extends to regions far into the conical nanopore (Fig. S9). The shape of the ionic current blockade is then mainly determined by the combined effects of geometrical exclusion of ions and ion concentration polarization. At a positive voltage, the ion concentration was depleted, so the effects of geometrical exclusion of ions and ion concentration polarization were cooperative, both decreasing the current. At a negative voltage, their effects were competitive, with the geometrical exclusion effect decreasing the current and ion concentration polarization effect increasing the current. The measured current is determined by the combination of these two effects. For instance, at -600 mV, the negative peak occurred at position e2 in Fig. 6 *A* and Fig. 6 *D* when the geometrical exclusion effect had the largest extent and the ion concentration enrichment was not so significant. The positive peak occurred before the end of the translocation - b2 in Fig. 6 *A* and Fig. 6 *D* coincident with significant ion concentration enrichment and small correction due to the geometrical exclusion. Fig. S10 shows a similar simulated current profile with a lower DNA charge density. If the salt concentration was low enough, the effect of ion concentration enhancement overtook the geometrical exclusion effect, resulting in a current increase only, as shown in Fig. 6 *B* and Fig. 6 *E*. The simulated event shapes for translocations in 100 mM KCl and 20 mM KCl solutions are very similar to the experimental results, as shown in Fig. 4 *A* and Fig. 4 *B*, Fig. S11 and Fig. 6 *E* respectively.

We also studied the effect of the pore geometry with 20 mM KCl solution by simulation (Fig. S12). The simulations showed that the ion concentration modulation was more significant for 12 nm tip diameters than that of 18 nm tip diameter. This result is expected due to a reduction in the relative importance of surface charges, in accordance with our experimental results. Overall our numerical simulations account for the variety of ionic current blockades observed in experiments over a large salt concentration range.

**CONCLUSIONS**

In summary, we have presented a comprehensive study of the direction, salt and voltage dependence of ionic current signatures for DNA translocations through conical asymmetric nanopores. At salt concentrations of ~4M, translocations in both directions caused current drop but with different amplitudes. With an intermediate salt concentration of ~1M, the current decreased during the forward translocations while it decreased at the beginning and increased at the end in a backward translocation event. At lower salt concentrations closer to physiological ionic strengths, only translocations from inside the pore to outside were observed due to the increasing influence of electroosmotic flow. Numerical simulations were conducted to explore the current change by solving the PNP-NS equations, and revealed that the ion concentration inside the pore was depleted at a positive voltage and enriched at a negative voltage. The direction-dependent ionic current blockades were found to be caused by differences in the effects of ion concentration modulation and geometrical exclusion of ions, with a cooperative action at a positive voltage and a competitive action at a negative voltage. Our results pave the way for optimizing biological molecule sensing with asymmetric conical nanopores by tuning voltage and salt concentrations.

## SUPPORTING MATERIAL

Supplementary figures and discussion are available in the supporting material.

## AUTHOR CONTRIBUTIONS

K. Chen, N. A. W. Bell, J. Kong and U. F. Keyser designed the research. K. Chen performed the nanopore measurements. N. A. W. Bell synthesized the DNA rulers. K. Chen., N. A. W. Bell, J. Kong, Y. Tian and U. F. Keyser wrote and edited the manuscript. U. F. Keyser oversaw the research.

## ACKNOWLEDGMENT
The authors thank K. Misiunas for discussion on COMSOL simulations. K. Chen acknowledges funding from Chinese Scholarship Council (201506210147). J. Kong acknowledges funding from Chinese Scholarship Council and Cambridge Trust. N. Bell and U. Keyser acknowledge funding from an ERC consolidator grant (Designerpores 647144).

## REFERENCES
1. Kasianowicz, J. J., Brandin, E., …, D. W. Deamer. 1996. Characterization of individual polynucleotide molecules using a membrane channel. *P. Natl. Acad. Sci. USA.* 93:13770-13773.

2. Storm, A. J., Chen, J. H., …, C. Dekker. 2003. Fabrication of solid-state nanopores with single-nanometre precision. *Nat. Mater.* 2:537-540.

3. Li, J., Gershow, M., …, J. A. Golovchenko. 2003. DNA molecules and configurations in a solid-state nanopore microscope. *Nat. Mater.* 2:611-615.

4. White, R. J., Ervin, E. N., …, H. S. White. 2007. Single ion-channel recordings using glass nanopore membranes. *J. Am. Chem. Soc.* 129:11766-11775.

5. Balan, A., Chien, C. C., …, M. Drndic. 2015. Suspended solid-state membranes on glass chips with sub 1-Pf capacitance for biomolecule sensing applications. *Sci. Rep.* 5, 17775.

6. Garaj, S., Hubbard, W., …, J. A. Golovchenko. 2010. Graphene as a subnanometre trans-electrode membrane. *Nature.* 467:190-193.

7. Liu, K., Feng, J., …, A. Radenovic. 2014. Atomically thin molybdenum disulfide nanopores with high sensitivity for DNA translocation. *ACS Nano.* 8:2504-2511.

8. Steinbock, L. J., Bulushev, R. D., …, A. Radenovic. 2013. DNA translocation through low-noise glass nanopores. *ACS Nano.* 7:11255-11262.

9. Li, W., Bell, N. A. W., …, U. F. Keyser. 2013. Single protein molecule detection by glass nanopores. *ACS Nano*. 7:4129-4134.

10. Tsutsui, M., Hongo, S., …, T. Kawai. 2012. Single-nanoparticle detection using a low-aspect-ratio pore. *ACS Nano*. 6:3499-3505.

11. Lan, W. J., Holden, D. A. …, H. S. White. 2011. Nanoparticle transport in conical-shaped nanopores. *Anal. Chem.* 83:3840-3847.

12. Coulter, W. H. Means for counting particles suspended in a fluid. 1953. U.S. Patent. 2656508, October 20.

13. Smeets, R. M. M., Keyser, U. F. …, C. Dekker. 2006. Salt dependence of ion transport and DNA translocation through solid-state nanopores. *Nano Lett.* 6:89-95.

14. Kowalczyk, S. W., and C. Dekker. 2012. Measurement of the docking time of a DNA molecule onto a solid-state nanopore. *Nano Lett.* 12:4159-4163.

15. Vlassarev, D. M., and J. A. Golovchenko. 2012. Trapping DNA near a solid-state nanopore. *Biophys. J.* 103:352-356.

16. Zanjani, M. B., Engelke, R. E., ..., M. Drndic. 2015. Up and down translocation events and electric double-layer formation inside solid-state nanopores. *Phys. Rev. E.* 92:022715.

17. Goyal, G., Freedman, K. J., and M. J. Kim. 2013. Gold Nanoparticle translocation dynamics and electrical detection of single particle diffusion using solid-state nanopores. *Anal. Chem.* 85:8180-8187.

18. Menestrina, J., Yang, C., …, Z. S. Siwy. 2014. Charged particles modulate local ionic concentrations and cause formation of positive peaks in resistive-pulse-based detection. *J. Phys. Chem. C.* 118:2391-2398.

19. Lan, W.-J., Kubeil, C., …, H. S. White. 2014. Effect of surface charge on the resistive pulse waveshape during particle translocation through glass nanopores. *J. Phys. Chem. C.* 118:2726-2734.

20. Cabello-Aguilar, S., Abou Chaaya, A., …, S. Balme. 2014. Experimental and simulation studies of unusual current blockade induced by translocation of small oxidized PEG through a single nanopore. *Phys. Chem. Chem. Phys.* 16:17883-17892.

21. Ivanov, A. P., Actis, P., …, J. B. Edel. 2015. On-demand delivery of single DNA molecules using nanopipets. *ACS Nano.* 9:3587-3595.

22. Schoch, R. B., Han, J., and P. Renaud. Transport phenomena in nanofluidics. 2008. *Rev. Mod. Phys.* 80:839-883.

23. Bell, N. A., Thacker, V. V., …, U. F. Keyser. 2013. Multiplexed ionic current sensing with glass nanopores. *Lab Chip.* 13:1859-1862.

24. Freedman, K. J., Otto, L. M., ..., J. B. Edel. 2016. Nanopore sensing at ultra-low concentrations using single-molecule dielectrophoretic trapping. *Nat. Commun.* 7.

25. Fraccari, R. L., Ciccarella, P., ..., T. Albrecht. 2016. High-speed detection of DNA translocation in nanopipettes. *Nanoscale*. 8:7604-7611.

26. Fraccari, R. L., Carminati, M., ..., T. Albrecht. 2016. High-bandwidth detection of short DNA in nanopipettes. *Faraday Discuss.*

27. Bell, N. A., and U. F. Keyser. 2015. Specific protein detection using designed DNA carriers and nanopores. *J. Am. Chem. Soc.* 137:2035-2041.

28. Bell, N. A., and U. F. Keyser. 2016. Digitally encoded DNA nanostructures for multiplexed, single-molecule protein sensing with nanopores. *Nat. Nanotechnol.*

29. Kong, J. Bell, N. A., and U. F. Keyser. 2016. Quantifying nanomolar protein concentrations using designed DNA carriers and solid-state nanopores. *Nano Lett.* 16:3557-3562.

30. Siwy, Z.S. Ion-current rectification in nanopores and nanotubes with broken symmetry. 2006. *Adv. Func. Mater.* 16:735-746.

31. Laohakunakorn, N., and U. F. Keyser. 2015. Electroosmotic flow rectification in conical nanopores. *Nanotechnology*. 26:275202.

32. Mihovilovic, M., Hagerty, N., and D. Stein. 2013. Statistics of DNA capture by a solid-state nanopore. *Phys. Rev. Lett.* 110:028102.

33. Steinbock, L. J., Lucas, A., ..., U. F. Keyser. 2012. Voltage-driven transport of ions and DNA through nanocapillaries. *Electrophoresis*. 33:3480-3487.

34. Qiu, Y., Vlassiouk, I., ..., Z. S. Siwy. 2016. Direction dependence of resistive-pulse amplitude in conically shaped mesopores. *Anal. Chem.* 88:4917-4925.

35. Kowalczyk, S. W., Wells, D. B., ..., C. Dekker. 2012. Slowing down DNA translocation through a nanopore in lithium chloride. *Nano Lett.* 12:1038-1044.

36. Hlushkou, D., Perry, J. M., ..., U. Tallarek. 2011. Propagating concentration polarization and ionic current rectification in a nanochannel-nanofunnel device. *Anal. Chem.* 84:267-274.

37. Smeets, R.M., Keyser, U.F., ..., C. Dekker. 2008. Noise in solid-state nanopores. *P. Natl. Acad. Sci. USA*. 105:417-421.

38. Bell, N. A.. and U. F. Keyser. Direct measurements reveal non-Markovian fluctuations of DNA threading through a solid-state nanopore. arXiv preprint arXiv:1607.04612.

39. Bell, N. A., Muthukumar, M., and Keyser, U. F. 2016. Translocation frequency of double-stranded DNA through a solid-state nanopore. *Phys. Rev. E*. 93:022401.

40. Chen, K., Shan, L., ..., Y. Tian. 2015. Biphasic resistive pulses and ion concentration modulation during particle translocation through cylindrical nanopores. *J. Phys. Chem. C* 119:8329-8335.

41. Keyser, U. F., Koeleman, B. N., ..., C. Dekker. 2006. Direct force measurements on DNA in a solid-state nanopore. *Nat. Phys.* 2:473-477.

42. Lu, B., Hoogerheide, D. P., …, D. Yu. 2012. Effective driving force applied on DNA inside a solid-state nanopore. *Phys. Rev. E.* 86:011921.

43. Schellman, J.A., and D. Stigter. 1977. Electrical double layer, zeta potential, and electrophoretic charge of double-stranded DNA. *Biopolymers.* 16:1415-1434.

44. Tsutsui, M., He, Y., …, T. Kawai. 2015. Particle trajectory-dependent ionic current blockade in low-aspect-ratio pores. *ACS Nano.* 10:803-809.

45. Kesselheim, S., Muller, W., and C. Holm. 2014. Origin of current blockades in nanopore translocation experiments. *Phys. Rev. Lett.* 112:018101.